\documentclass{article}
\usepackage{Adam}
\usepackage{enumitem} 
\usepackage{ifthen} 
\usepackage{mathrsfs} 
\usepackage{textcase} 

\usepackage{fancyheadings} 

\usepackage{hyphenat} 

\newcommand{\langfont}[1]{\textsc{#1}}
\newcommand{\wherePoly}[2]{where $#1 \geqslant 1/\poly(#2)$}
\newcommand{\seeglos}{\footnote{definition provided in the glossary (appendix \ref{glossary})}}

\title{\textbf{QMA-complete problems}}
\date{August 22, 2013}
\author{Adam D. Bookatz\thanks{Center for Theoretical Physics, Massachusetts Institute of Technology, Cambridge, MA, USA; \texttt{bookatz@mit.edu}}}
\begin{document}
\maketitle
\thispagestyle{fancy} 
\abstract{
In this paper we give an overview of the quantum computational complexity class QMA and a description of known QMA-complete problems to date\footnote{The reader is invited to share more proven QMA-complete problems with the author.}. Such problems are believed to be difficult to solve, even with a quantum computer, but have the property that if a purported solution to the problem is given, a quantum computer would easily be able to verify whether it is correct.
An attempt has been made to make this paper as self-contained as possible so that it can be accessible to computer scientists, physicists, mathematicians, and quantum chemists. Problems of interest to all of these professions can be found here.
}

%
%
\section{Introduction}
\subsection{Background}
The class QMA is the natural extension of the classical class NP to the quantum computing world. NP is an extremely important class in classical complexity theory, containing (by definition) those problems that 
have a short proof (or witness) that can be efficiently checked to verify that a valid solution to the problem exists. The class NP is of great importance because many interesting and important problems have this property -- they may be difficult to solve, but given a solution, it is easy to verify that the solution is correct. 

The probabilistic extension of NP is the class MA, standing for ``Merlin-Arthur". Unlike in NP where witnesses must be verifiable with certainty, in MA valid witnesses need only be accepted with probability greater than $2/3$ (and invalid witnesses rejected with probability greater than $2/3$). MA can be thought of as the class of languages $L$ for which a computationally-unbounded but untrustworthy prover, Merlin, can convince (with high probability) a verifier, Arthur (who is limited to polynomial-time computation), that a particular instance $x$ is in $L$. Furthermore, when the instance $x$ is not in $L$, the probability of Merlin successfully cheating must be low.

Because quantum computers are probabilistic by nature (as the outcome of a quantum measurement can generally be predicted only probabilistically),
the \emph{natural} quantum analogue of NP is actually the quantum analogue of MA, whence the quantum class QMA -- Quantum-Merlin-Arthur\footnote{Initially QMA was referred to as BQNP \cite{KSV02}; the name QMA was coined in \cite{Watrous00}.}. 
QMA, then, is the class of languages for which small \emph{quantum} witness states exist that enable one to prove, with high probability, whether a given string belongs to the language by whether the witness state causes an efficient \emph{quantum verifier circuit} to output 1. It was first studied by Kitaev\cite{KSV02} and Knill\cite{Knill96}. A more precise definition will be given later. 
 
The history of QMA takes its lead from the history of NP.
In complexity theory, one of the most important results about NP was the first proof that it contains 
complete problems.
A problem is NP-hard if, given the ability to solve it, one can also efficiently (that is, with only polynomial overhead) solve \emph{any} other NP problem; in other words, a problem is NP-hard if any NP problem can be \emph{reduced} to it. If, in addition to being NP-hard, a problem is itself in NP, it is called an NP-complete problem, and can be considered among the hardest of all the problems in NP.
Two simple examples of NP-complete problems are \langfont{Boolean satisfiability} (SAT) and \langfont{circuit satisfiability} (CSAT). 
The problem CSAT is to determine, given a Boolean circuit, whether there exists an input that will be evaluated by the circuit as true. The problem SAT is to determine, given a set of clauses containing Boolean variables, whether there is an assignment of those variables that will satisfy all of the clauses.  If the clauses are restricted to containing at most $k$ literals each, the problem is called $k$-SAT. One may also consider the problem MAX-SAT, which is to determine, given a set of clauses (of Boolean variables) and an integer $m$, whether at least $m$ clauses can be simultaneously satisfied. 

The fact that a complete problem exists for NP is actually trivial, 
as the problem of deciding whether there exists a (small) input that will be accepted (in polynomial time) by a given Turing machine is trivially NP-complete; rather, the importance of NP-completeness is due to the existence of interesting NP-complete problems. The famous Cook-Levin theorem, which pioneered the study of NP-completeness, states that SAT 
is NP-complete. A common way of proving this theorem is to first show that the above trivial NP-complete problem can be reduced to CSAT, 
 and to then show that CSAT can be reduced to SAT. 
 
The quantum analogue of CSAT is the \langfont{quantum circuit satisfiability} problem (QCSAT), which is trivially QMA-complete (since QMA is defined in terms of quantum circuits). But QMA was found to have other, natural complete problems too. The most important of these, the \langfont{$k$-local Hamiltonian } problem \cite{KSV02}, was defined by Kitaev \cite{KSV02} inspired by Feynman's ideas on Hamiltonian quantum computing \cite{Feynman82}. This problem is a quantum analogue of MAX-SAT, in which Boolean variables are replaced by qubits and clauses are replaced by local Hamiltonians (which may be viewed as local constraints on the qubits); it is defined formally below in \ref{k-local}.
Just as the Cook-Levin theorem opened the study of NP-completeness by showing that SAT is NP-complete, so too the study of QMA-completeness began by showing that \langfont{5-local Hamiltonian} is QMA-complete.

However, unlike NP, for which thousands of complete problems are known, there are currently relatively few known QMA-complete problems. In this paper we will survey many, if not all, of them. This paper divides the QMA-complete problems into three main groups and one subgroup:
\begin{itemize}
\item Quantum circuit/channel property verification (V)
\item Hamiltonian ground state estimation (H)
	\begin{itemize}
	\item Quantum $k$-SAT (S)
	\end{itemize}
\item Density matrix consistency (C)
\end{itemize}
The letters in parentheses are used as labels to identify the group.

\subsection{Formal definition of QMA}
We now give a formal definition of QMA.

\begin{definition}
QMA is the set of all languages $L\subset\{0,1\}^*$ for which there exists a (uniform family of) 
quantum polynomial-time verifier circuit $V$ such that for every $x\in \{0,1\}^*$ of length $n=|x|$,
\begin{description}[leftmargin=80pt, align=right, font=\normalfont, labelindent=60pt, itemsep=1ex]
\item [if $x\in L$] then there exists a poly$(n)$-qubit witness state $\ket{\psi_x}$ such that $V\big(x,\ket{\psi_x}\big)$ accepts with probability $\geqslant 2/3$
\item [if $x\not\in L$] then for every purported poly$(n)$-qubit witness state $\ketpsi$, $V\big(x,\ketpsi\big)$ accepts with \\probability~$\leqslant 1/3$.
\end{description}
\end{definition}
  
Although the definition above used the numbers $2/3$ and $1/3$ (as is standard), 
we can generally define the class QMA$(b,a)$: Given functions $a,b : \N \rightarrow (0,1)$ with $b(n)-a(n)\geqslant 1/\poly(n)$, a language is in QMA$(b,a)$ if $2/3$ and $1/3$ in the definition above are replaced by $b$ and $a$, respectively. 
It is important to note that doing this does not change the class: QMA$(2/3,1/3)$ = QMA$(1-\epsilon, \epsilon)$ provided that $\epsilon \geqslant 2^{-\poly(n)}$. Moreover, in going from QMA$(2/3,1/3)$ to QMA$(1-\epsilon, \epsilon)$, the amplification procedure can be carried out in such a way that the same witness is used, i.e. Merlin need only ever send a single copy of the witness state. \cite{MW05}

When $b=1$, i.e. when the witness must be verifiable with no error, the class is called QMA$_1$; thus QMA$_1$ = QMA$(1,1/3)$ = QMA$(1,\epsilon)$. For the classical complexity class MA it is known that MA = MA$_1$, 
and also for the class QCMA, which is QMA restricted to classical witnesses, it has recently been shown~\cite{JKNN12} that QCMA=QCMA$_1$,
but it is still an open question whether QMA=QMA$_1$. Several QMA$_1$-complete problems are presented in this paper.

Furthermore, it should be noted that QMA actually consists of promise problems, meaning that when considering whether Merlin can truthfully convince Arthur or trick Arthur, we restrict our consideration to a subset of possible instances -- we may assume that we are promised that our instance of consideration falls in this subset. In physical problems, this restriction could correspond to a limitation in the measurement precision available to us. 
With the above remarks, we can write the definition of QMA in a style matching that of the problem definitions provided in this paper.
\begin{definition}[QMA]
A promise problem $L = L_{\text{yes}}\cup L_{\text{no}}\subset\{0,1\}^*$ is in QMA if there exists a quantum polynomial-time verifier circuit $V$ such that for every $x\in \{0,1\}^*$ of length $n=|x|$,
\begin{description}[leftmargin=80pt, align=right, font=\normalfont, labelindent=60pt, itemsep=1ex]
\item [(yes case)] if $x\in L_{\text{yes}}$ then $\exists \poly(n)$-qubit state $\ket{\psi_x}$ such that $\Pr\Big[V\big(x,\ket{\psi_x}\big) \text{ accepts }\Big] \geqslant b$
\item [(no case)] if $x\in L_{\text{no}}$ then $\forall \poly(n)$-qubit states $\ketpsi$, $\Pr\Big[V\big(x,\ketpsi\big) \text{ accepts }\Big] \leqslant a$
\end{description}
promised that one of these is the case (i.e. either $x$ is in $L_{\text{yes}}$ or $L_{\text{no}}$), \\
{\wherePoly{b-a}{n}} and $0<\epsilon<a<b<1-\epsilon$, with $\epsilon \geqslant 2^{-\poly(n)}$. If, instead, the above is true with $b=1$, then $L$ is in the class QMA$_1$.
\end{definition}

Except for a glossary at the end, which provides the definitions of several basic reoccurring mathematical terms that appear in this work, the remainder of this paper is devoted to listing known QMA-complete problems, along with their description and sometimes a brief discussion. 
When a problem is given matrices, vectors, or constants as inputs, it is assumed that they are given to precision of $\poly(n)$ bits. When a problem is given a unitary or quantum circuit, $U_x$, it is assumed that the problem is actually given a classical description $x$ of the corresponding quantum circuit, which consists of $\poly(|x|)$ elementary gates. Likewise, quantum channels are specified by efficient classical descriptions.

This paper has attempted to be as self-contained as possible, but for a more complete description and motivation of a problem, the reader is invited to consult the references provided. An attempt has been made to include as many currently known QMA-complete and QMA$_1$-complete problems as possible, but it is, of course, unlikely that this goal has been accomplished in full. The reader is invited to share other proven QMA-complete problems with the author for their inclusion in future versions of this work. Note that this paper has restricted itself to QMA-complete and QMA$_1$-complete problems; it does not include other QMA-inspired classes, such as QMA(2) (when there are multiple unentangled Merlins) or QCMA (when the witness is classical).

%
%
\setlist[enumerate,1]{itemsep=20pt}
\newcommand{\setEnumLabel}[1]{\setlist[enumerate,1]{label*=\fbox{{#1}-\arabic*} , ref={#1}-\arabic*  ,itemsep=20pt    }}
\newcommand{\EnumItem}[1]{\item{\textbf{\MakeTextUppercase{#1}}} \\} 
%

%
%
\newcommand{\probIntro}[1]{#1 \vspace{6pt}}
\newcommand{\probPreamble}[1]{\textit{Preliminary information}:\\ #1 \vspace{6pt}}
\newcommand{\probDesc}[4]{
\fbox{
\parbox{400pt}{
Problem: #1, \\
determine whether:
\begin{description}[leftmargin=80pt, align=right, font=\normalfont, labelindent=60pt, itemsep=1ex]
\item [(yes case)] #2     
\item [(no case)] #3,
\end{description}
promised one of these to be the case\ifthenelse{\equal{#4}{}}{.}{,\\#4.}
}}
}
\newcommand{\probTheorem}[3]{Theorem: QMA-complete #1 [proven by #2 \cite{#3}]}      
\newcommand{\probTheoremO}[3]{Theorem: QMA$_1$-complete #1 [proven by #2 \cite{#3}]} 
\newcommand{\probTheoremH}[3]{Theorem: QMA-hard #1 [proven by #2 \cite{#3}]} 
\newcommand{\probRed}[2]{Hardness reduction from: #1 (#2)}
\newcommand{\probClassical}[1]{Classical analogue: #1.}
\newcommand{\probNotes}[1]{\begin{description}[font=\normalfont\emph, leftmargin=31pt]
\item[Notes:] #1
\end{description}}

\newcommand{\gap}{\vspace{10 mm}}

%
%
\section{Quantum circuit/channel property verification}
\setEnumLabel{V} \begin{enumerate}

\EnumItem{Quantum Circuit-SAT (QCSAT) \label{QCSAT}} 
\probDesc{
Given a quantum circuit $V$ on $n$ witness qubits and $m=\poly(n)$ ancilla qubits}
{$\exists$ $n$-qubit state $\ket{\psi}$ such that $V(\ket{\psi}\ket{0\ldots 0})$ accepts with probability $\geqslant b$, i.e. outputs a state with $\ket{1}$ on the first qubit with amplitude-squared $\geqslant b$}
{$\forall$ $n$-qubit state  $\ket{\psi}$, $V(\ket{\psi}\ket{0\ldots 0})$ accepts with probability $\leqslant a$}
{\wherePoly{b-a}{n} and $\ket{0\ldots 0}$ is the all-zero $m$-qubit ancilla state}

This problem is QMA-complete immediately from the definition of QMA.

\EnumItem{non-identity check \label{non-identity}}
\probDesc{
Given a unitary $U$ implemented by a quantum circuit on $n$ qubits, 
determine whether $U$ is \textit{not} close to a trivial unitary (the identity times a phase), i.e.}
{$\forall \phi\in[0,2\pi), \norm{U - e^{i \phi}\id} \geqslant b$}
{$\exists \phi\in[0,2\pi)$ such that $\norm{U - e^{i \phi}\id} \leqslant a$}
{\wherePoly{b-a}{n}}

\probTheorem{}{Janzing, Wocjan, and Beth}{JWB05}\\
\probTheorem{even for small-depth quantum circuits}{Ji and Wu}{JW09}\\
\probRed{QCSAT}{\ref{QCSAT}}

\EnumItem{non-equivalence check}
\probIntro{This problem, a generalisation of \langfont{non-identity check} (\ref{non-identity}), is to determine whether two quantum circuits (do not) define approximately the same unitary (up to phase) on some chosen invariant subspace. The subspace could, of course, be chosen to be the entire space, but in many cases one is interested in restricting their attention to a subspace, e.g. one defined by a quantum error-correcting code.
}

\probDesc{
Given two unitaries, $U_1$ and $U_2$, implemented by a quantum circuit on $n$ qubits, let $\mathcal{V}$ be a common invariant subspace of $(\C^2)^{\otimes n}$ specified by a quantum circuit $V$ (that ascertains with certainty whether a given input is in $\mathcal{V}$ or not). The problem is to determine, given $U_1$, $U_2$, and $V$, whether the restrictions of $U_1$ and $U_2$ to $\mathcal{V}$ are not approximately equivalent, i.e.}
{$\exists \ \ket{\psi}\in\mathcal{V}$ such that $\forall \phi\in[0,2\pi), \norm{(U_1 U_2^\dag - e^{i \phi}\id)\ketpsi} \geqslant b$}
{$\exists \ \phi\in[0,2\pi)$ such that $\forall \ket{\psi}\in\mathcal{V},  \norm{(U_1 U_2^\dag - e^{i \phi}\id)\ketpsi} \leqslant a$}
{\wherePoly{b-a}{n}}

\probTheorem{}{Janzing, Wocjan, and Beth}{JWB05}\\
\probRed{\langfont{non-identity check}}{\ref{non-identity}}

\EnumItem{Mixed-state non-identity check \label{mixed-non-identity}}
\probIntro{In this problem, either the given circuit acts like some unitary $U$ that is far from the identity, or else it acts like the identity. This is very similar to \langfont{non-identity check} (\ref{non-identity}), but allows mixed-state circuit inputs. The diamond norm used here is defined in the glossary (appendix \ref{glossary}).}

\probDesc{
Given a quantum circuit $C$ on $n$-qubit density matrices}
{$\Vert C-\id \Vert_{\diamondsuit} \geqslant 2-\epsilon$ and there is an efficiently implementable unitary $U$ and state $\ketpsi$ such that $\Vert C(\selfketbra{\psi}) - U\selfketbra{\psi}U^\dag\Vert_{tr} \leqslant \epsilon$ and $\Vert U\selfketbra{\psi}U^\dag - \selfketbra{\psi}\Vert_{tr} \geqslant 2-\epsilon$}
{$\Vert C-\id \Vert_{\diamondsuit} \leqslant\epsilon$}
{where $1 > \epsilon \geqslant 2^{-\poly(n)}$}

\probTheorem{}{Rosgen}{Rosgen11b}\\
\probRed{Quantum circuit testing (see appendix~\ref{app:hard})}{\ref{circuit_testing}}

\EnumItem{non-isometry testing \label{non-isometry}} 
\probPreamble{
This problem tests to see if a quantum channel is not almost a linear isometry (given a mixed-state quantum circuit description of the channel).
\begin{definition}[isometry]
A linear isometry is a linear map $U : {\cal H}_1 \rightarrow {\cal H}_2$ that preserves inner products, i.e. $U^\dag U = \id_{{\cal H}_1}$.
\end{definition}
Note that this is more general than a unitary operator, as ${\cal H}_1$ and ${\cal H}_2$ may have different sizes and $U$ need not be surjective. 
Practically speaking, isometries are the operations involving unitaries that have access to fixed ancillae (say, ancillae starting in the $\ket{0}$ state). 
%
%
This problem asks how far from an isometry the input is, so it requires a notion of approximate isometries. A characterising property of isometries is that they map pure states to pure states, even in the presence of a reference system; therefore, the notion of an approximate isometry is defined in terms of how mixed the output of a channel is in the presence of a reference system.
\begin{definition}[$\epsilon$-isometry]
A quantum channel $\Phi$ that is a linear transformation from ${\cal H}_1$ to ${\cal H}_2$ is an $\epsilon$-isometry if $\forall \ket{\psi} \in {\cal H}_1\otimes {\cal H}_1$, we have $\norm{ (\Phi\otimes\id_{{\cal H}_1})(\selfketbra{\psi})} \geqslant 1-\epsilon$. i.e. it maps pure states (in a combined input and reference system) to almost-pure states. The norm appearing in this definition is the operator norm\seeglos.
\end{definition}}

\probDesc{
Given a quantum channel $\Phi$ that takes density matrices of ${\cal H}_1$ to density matrices of ${\cal H}_2$}
{$\Phi$ is not an $\epsilon$-isometry, i.e. $\exists \ket{\psi}$ such that $\norm{ (\Phi\otimes\id_{{\cal H}_1})(\selfketbra{\psi})} \leqslant\epsilon$}
{$\Phi$ is an $\epsilon$-isometry, i.e. $\forall\ket{\psi}$, $\norm{ (\Phi\otimes\id_{{\cal H}_1})(\selfketbra{\psi})} \geqslant 1-\epsilon$}
{}

\probTheorem{when $0<\epsilon < 1/19$}{Rosgen}{Rosgen11a,Rosgen11b} 

\probRed{QCSAT}{\ref{QCSAT}} 

\EnumItem{Detecting insecure quantum encryption \label{insecure_q_enc}}
\probIntro{In this problem, we wish to determine whether the given purported encryption channel $E$ is insecure on a large subspace (for any key), or is close to being perfectly secure. The diamond norm used here is defined in the glossary (appendix~\ref{glossary}).}

\probPreamble{
A private channel is a quantum channel with a classical key such that the input state cannot be determined from the output state without the key. Formally, it is defined as follows.
\begin{definition} [$\epsilon$-private channel] Suppose $E$ is a channel taking as input an integer $k\in\lbrace 1,.\ldots,K\rbrace$ and quantum state in space ${\cal H}_1$, and producing a quantum output in space ${\cal H}_2$, with $\text{dim\,}{\cal\,H}_1\leqslant\text{dim }{\cal H}_2$. Let $E_k$ be the quantum channel where the integer input is fixed as $k$. Let $\Omega$ be the completely depolarizing channel that maps all density matrices to the maximally mixed state.
 $E$ is an $\epsilon$-private channel if $\Vert \frac{1}{K} \sum_k E_k - \Omega \Vert_{\diamondsuit} \leqslant \epsilon$ (so if the key $k$ is not known, the output of $E$ gives almost no information about the input)
 and
 there is a polysize decryption channel $D$ (operating on the same space as $E$) such that $\forall k, \Vert D_k \circ E_k -\id\Vert_{\diamondsuit} \leqslant \epsilon$ (i.e. if $k$ is known, the output can be reversed to obtain the input).
 \end{definition}
}

\probDesc{
Let $\delta\in(0,1]$. Given circuit $E$, which upon input $k$ implements channel $E_k$ acting from space ${\cal H}_1$ to ${\cal H}_2$ (with $\text{dim\,}{\cal H}_1\leqslant\text{dim\,}{\cal H}_2$)}
{$\exists$ subspace $S$, with $\text{dim\,} S \geqslant (\text{dim\,} {\cal H}_1)^{1-\delta}$, such that for any $k$ and any reference space $\cal R$, if $\rho$ is a density matrix on $S\otimes \cal R$ then $\Vert (E_k\otimes\id_{R})(\rho) - \rho\Vert_{tr} \leqslant \epsilon$}
{$E$ is an $\epsilon$-private channel}
{where $1 > \epsilon \geqslant 2^{-\poly}$}

\probTheorem{for $0<\epsilon<1/8$}{Rosgen}{Rosgen11b}\\
\probRed{Quantum circuit testing (see appendix~\ref{app:hard})}{\ref{circuit_testing}}

\probNotes{In this problem, channels are given as mixed-state circuits.}

\EnumItem{Quantum clique}
\probIntro{
This is the quantum analogue of the NP-complete problem \langfont{largest independent set} on a graph $G$, which asks for the size of the largest set of vertices in which no two vertices are adjacent. According to the analogy, the graph $G$ becomes a channel, and two inputs are `adjacent' if they can be confused after passing through the channel, i.e. if there is an output state that could have come from either of the two input states. In this quantum QMA-complete problem, the channel is a quantum entanglement-breaking channel $\Phi$ and the problem is to find the size of the largest set of input states that cannot be confused after passing through the channel, that is, to determine if there are $k$ inputs $\rho_1,\ldots,\rho_k$ such that $\Phi(\rho_1),\ldots,\Phi(\rho_k)$ are (almost) orthogonal under the trace inner product. Regarding the name, note that the NP-complete problems \langfont{largest independent set} and \langfont{largest clique} (which asks for the largest set of vertices, all of which are adjacent) are essentially the same: a set of vertices is an independent set on a graph $G$ if and only if it is a clique on the complement of $G$.
}

\probPreamble{
Let $S$ be the SWAP gate, so $S\ket{\psi}\otimes\ket{\phi} = \ket{\phi}\otimes\ket{\psi}$. Note that $\trP(\sigma_1\sigma_2)=\trP(S\, \sigma_1\otimes\sigma_2)$ for all density matrices $\sigma_1$ and $\sigma_2$, so the right hand side can be used to evaluate the trace inner product (and therefore determine orthogonality) of $\sigma_1$ with $\sigma_2$.
For any density matrix $\rho$ on $k$ registers, let $\rho^i$ denote the result of tracing out all but the $i$th register of $\rho$. Similarly, define $\rho^{i,j} = \trP_{\lbrace 1,\ldots k\rbrace \smallsetminus{\lbrace i,j \rbrace}} (\rho)$.

\begin{definition}[entanglement-breaking channel; q-c channel]
A quantum channel $\Phi$ is \textit{entanglement-breaking} if there are POVM (Hermitian, positive-semidefinite operators that sum to the identity) $\lbrace M_i \rbrace$ and states $\sigma_i$ such that $\Phi(\chi) = \sum_i \trP(M_i\chi)\sigma_i$. In this case it is a fact that $\Phi^{\otimes 2} (\rho^{1,2})$ is always a separable state. 
If the $\sigma_i$ in the above definition can be chosen to be $\sigma_i = \selfketbra{i}$, where $\ket{i}$ are orthogonal states, then $\Phi$ is called a \text{quantum classical channel} (q-c channel).
\end{definition}}

\probDesc{
Given an integer $k$ and a quantum entanglement-breaking channel $\Phi$ acting on $n$-qubit states}
{$\exists\ \rho_1\otimes\cdots\otimes\rho_k$ such that $\sum_{i,j} \trP(S \Phi(\rho_i)\otimes\Phi(\rho_j)) \leqslant a$}
{$\forall\ k$-register state $\rho$, $\sum_{i,j} \trP(S \Phi^{\otimes 2}(\rho^{i,j})) \geqslant b$}
{where $b$ and $a$ are inverse-polynomially separated}

There are two theorems associated with this problem.
\begin{enumerate}
\item \probTheorem{}{Beigi and Shor}{BS07}\\ 
\item \probTheoremO{when $a=0$ and $\Phi$ is further restricted to q-c channels}{Beigi and Shor}{BS07}
\end{enumerate}
\probRed{\langfont{$k$-local Hamiltonian}}{\ref{k-local}}\\
\probClassical{\langfont{largest independent set} is NP-complete}

\EnumItem{quantum non-expander}
\probIntro{
A quantum expander is a superoperator that rapidly takes density matrices towards the maximally mixed state. The \langfont{quantum non-expander} problem is to check whether a given superoperator is \textit{not} a good quantum expander. This problem uses the Frobenius norm\seeglos.
}

\probPreamble{
A density matrix can always be written as $\rho = I + A$, where $I$ is the maximally mixed state and $A$ is traceless. A quantum expander is linear (and unital), so $\Phi(\rho) = I + \Phi(A)$, which differs from $I$ by $\Phi(A)$. Thus a good quantum expander rapidly kills traceless matrices. We have the following formal definition.
\begin{definition}[quantum expander]
Let $\Phi$ be a superoperator acting on $n$-qubit density matrices and obeying $\Phi(\rho) = \frac{1}{D}\sum_d U_d \rho U_d^\dag$ where $\lbrace U_d : d=1,\ldots D \rbrace$ is a collection of $D=\poly(n)$ efficiently-implementable unitary operators. $\Phi$ is a $\kappa$-contractive quantum expander if $\forall\ 2^n\times 2^n$ traceless matrix~$A$, $\norm{\Phi(A)}_F \leqslant \kappa \norm{A}_F$.
\end{definition} 
}

\probDesc{Given a superoperator $\Phi$ that can be written in the form appearing in the above definition}
{$\Phi$ is not a $b$-contractive quantum expander}
{$\Phi$ is an $a$-contractive quantum expander}
{\wherePoly{b-a}{n}}

\probTheorem{}{Bookatz, Jordan, Liu, and Wocjan}{BJLW12}\\  
\probRed{QCSAT}{\ref{QCSAT}}

\end{enumerate} 
%
%
\section{Hamiltonian ground-state energy estimation}
\setEnumLabel{H} \begin{enumerate}

\EnumItem{$k$-local Hamiltonian \label{k-local}} 
\probIntro{
This is the problem of estimating the ground-state energy\footnote{see the glossary (appendix \ref{glossary}) for a very brief definition of these terms} 
of a Hamiltonian in which all interactions are $k$-local, that is, they only ever involve at most $k$ particles at a time. Formally, $H$ is a $k$-local Hamiltonian if $H=\sum_i H_i$ where each $H_i$ is a Hermitian operator acting (non-trivially) on at most $k$ qubits. 
In addition to restricting the locality of a Hamiltonian in terms of the number of qubits on which it acts, one can also consider geometric restrictions on the Hamiltonian. Indeed, one can imagine a 2-local Hamiltonian in which interactions can only occur between neighbouring sites, e.g. $H = \sum_{i=1}^{n-1} H_{i,i+1}$ where each $H_{i,i+1}$ acts non-trivially only on particles $i$ and $i+1$ arranged on a line. The results of these considerations will also be mentioned below. Note that all these problems use the operator norm\addtocounter{footnote}{-1}\addtocounter{Hfootnote}{-1}\footnotemark}.

\probDesc{
Given a $k$-local Hamiltonian on $n$ qubits, $H=\sum_{i=1}^r H_i$, where $r=\poly(n)$ and each $H_i$ acts non-trivially on at most $k$ qubits and has bounded operator norm $\norm{H_i}\leqslant \poly(n)$} 
{$H$ has an eigenvalue less than $a$}
{all of the eigenvalues of $H$ are larger than $b$}
{\wherePoly{b-a}{n}}

\probTheorem{for $k\geqslant 2$}{Kempe, Kitaev, and Regev}{KKR06}\\
\probRed{QCSAT}{\ref{QCSAT}}\\

Additionally, it has been proved that it is:
\begin{enumerate}
\item \probTheorem{when $k=O(\log n)$ (still provided $k\geqslant 2$)}{Kitaev}{KSV02}\\
\item \probTheorem{even when $k=3$ with constant norms, i.e. $\norm{H_i}=O(1)$}{Nagaj}{NM07}\\
\item \probTheorem{even when 2-local on a line of 8-dimensional qudits\addtocounter{footnote}{-1}\addtocounter{Hfootnote}{-1}\footnotemark, i.e. when the qudits are arranged on a line and only nearest-neighbour interactions are present}{Hallgren, Nagaj, and Narayanaswami\footnote{improving the work by Aharonov, Gottesman, Irani, and Kempe\cite{AGIK09} who showed this for 12-dimensional qudits}}{HNN13} \\
\item \label{2D-lattice} \probTheorem{even when 2-local on a 2-D lattice}{Oliveira and Terhal}{OT08} \\
\item \probTheorem{even for interacting bosons under two-body interactions}{Wei, Mosca, and Nayak}{WMN10} \\
\item \probTheorem{even for interacting fermions under two-body interactions}{Whitfield, Love, and Aspuru-Guzik}{WLA12} \\
\item \label{realHam} \probTheorem{even when restricted to real 2-local Hamiltonians}{Biamonte and Love}{BL08}\\
\item \label{stochastic_ham} \probTheorem{even for stochastic\seeglos\ Hamiltonians (i.e. symmetric Markov matrices) when $k\geqslant 3$}{Jordan, Gosset, and Love}{JGL10}\\
\end{enumerate}

\probNotes{
For $k=1$, the \langfont{1-local Hamiltonian} is in P \cite{KKR06}. \\
Many other simple modifications of \langfont{$k$-local Hamiltonian} are also QMA-complete. For example\footnote{This result is from \cite{GK12}, who actually defined their problem for finding the highest energy of a positive-semidefinite Hamiltonian. Their interest lay in finding approximation algorithms for this problem.}, QMA-completeness is not changed when restricting to dense $k$-local Hamiltonians, i.e. for a negative-semidefinite Hamiltonian when the ground energy is (in absolute value) $\Omega(n^k)$.}

\probClassical{MAX-$k$-SAT is NP-complete for $k\geqslant 2$.\\
This problem may easily be rephrased in terms of satisfying constraints imposed by the $H_i$ terms. The yes case corresponds to the existence of a state that violates, in expectation value, only fewer than $a$ weighted-constraints; the no case, to all states violating, in expectation value, at least $b$ weighted-constraints. This problem can therefore be viewed as estimating the largest number of simultaneously satisfiable constraints, whence the analogy to MAX-SAT}

\EnumItem{excited $k$-local Hamiltonian}
\probIntro{
We have seen that estimating the ground-state energy of a Hamiltonian is QMA-complete. The current problem shows that estimating the low-lying excited energies of a Hamiltonian is QMA-complete; specifically, estimating the $c^\text{th}$ energy eigenvalue of a $k$-local Hamiltonian is QMA-complete when $c=O(1)$.
}

\probDesc{
Given a $k$-local Hamiltonian $H$ on $n$ qubits} 
{the $c^\text{th}$ eigenvalue of $H$ is $\leqslant a$}
{the $c^\text{th}$ eigenvalue of $H$ is $\geqslant b$}
{\wherePoly{b-a}{n}}

\probTheorem{for $c=O(1)$ and $k\geqslant 3$}{Jordan, Gosset, and Love}{JGL10} \\
\probRed{\langfont{2-local Hamiltonian}}{\ref{k-local}}

\EnumItem{highest energy of a $k$-local stoquastic Hamiltonian}
\probIntro{Problem \ref{stochastic_ham} states that finding the lowest eigenvalue of a stochastic\seeglos\addtocounter{footnote}{-1}\addtocounter{Hfootnote}{-1}
Hamiltonian is QMA-complete. Since if $H$ is a stochastic Hamiltonian then $-H$ is stoquastic\footnotemark, we also have QMA-completeness for the problem of estimating the largest energy of a stoquastic Hamiltonian.} 

\probDesc{
Given a $k$-local stoquastic Hamiltonian $H$ on $n$ qubits}
{$H$ has an eigenvalue greater than $b$}
{all of the eigenvalues of $H$ are less than $a$}
{\wherePoly{b-a}{n}}

\probTheorem{for $k\geqslant 3$}{Jordan, Gosset, and Love}{JGL10}\\
\probRed{\langfont{$k$-local stochastic Hamiltonian} (\ref{stochastic_ham}) which itself is from}{\ref{BL08-1}}

\probNotes{\langfont{$k$-local stoquastic Hamiltonian}, i.e. finding the lowest energy rather than the highest energy, is in AM. \cite{BDOT08}}

\EnumItem{Separable $k$-local Hamiltonian}
\probIntro{
This problem is the \langfont{$k$-local Hamiltonian} problem with the extra restriction that the quantum state of interest be a separable state, i.e. the question is whether there is a \textit{separable} state with energy less than $a$ (or greater than $b$). Separable, here, is with respect to a given partition of the space into two sets, between which the state must not be entangled.

}

\probDesc{
Given the same input as described in the \langfont{$k$-local Hamiltonian} problem, as well as a partition of the qubits to disjoint sets $\calA$ and $\calB$}
{$\exists \ketpsi = \ketpsi_A\otimes\ketpsi_B$, with $\ketpsi_A\in\calA$ and $\ketpsi_B\in\calB$, such that $\bra{\psi}H\ket{\psi}\leqslant a$}
{$\forall \ketpsi = \ketpsi_A\otimes\ketpsi_B$, with $\ketpsi_A\in\calA$ and $\ketpsi_B\in\calB$, $\bra{\psi}H\ket{\psi}\geqslant b$}
{\wherePoly{b-a}{n}}

\probTheorem{}{Chailloux and Sattath}{CS12}\\
\probRed{\langfont{$k$-local Hamiltonian}}{\ref{k-local}}

\probNotes{
Interestingly, although the QMA-hardness proof follows immediately from \langfont{$k$-local Hamiltonian}, the ``in QMA" proof is non-trivial and relies on the \langfont{local consistency problem} (\ref{local_consistency}).
}


\EnumItem{Physically relevant Hamiltonians}
\langfont{2-local Hamiltonian} is also QMA-complete when the Hamiltonian is restricted to various physically-relevant forms. These Hamiltonians may serve as good models for phenomena found in nature, or may be relatively easy to physically implement.

We will not explain all of the relevant physics and quantum chemistry here. However, we use the following notations:

The Pauli matrices $X$, $Y$, and $Z$ are denoted as
\[
\sigma^x = X, \quad\quad \sigma^y = Y, \quad\quad  \sigma^z = Z, \quad\quad \boldsymbol{\sigma} = (\sigma^x,\sigma^y,\sigma^z) .
\]
When particles are on a lattice,
$\langle i,j \rangle$ denotes nearest neighbours on the lattice.
%
An electron on a lattice is located at some lattice site $i$ 
 and can be either spin-up ($\uparrow$) or spin-down ($\downarrow$). 
The operators $a^\dag_{i,s}$ and $a_{i,s}$ are the fermionic raising and lowering operators, respectively; they create and annihilate an electron of spin $s\in\lbrace\uparrow,\downarrow\rbrace$ at site 
 $i$, respectively.
The operator corresponding to the number of electrons of spin $s$ at site $i$ is $n_{i,s} = a^\dag_{i,s} a_{i,s}$.

Note that proving the QMA-completeness of physical Hamiltonians is related to the goal of implementing adiabatic quantum computation: techniques used to prove that a Hamiltonian is QMA-complete are often also used to prove that it is universal for adiabatic quantum computation.

\begin{enumerate}


\item \label{BL08-1} The 2-local Hamiltonian
\[
H_{ZZXX} = \sum_i h_i \sigma^z_i + \sum_i d_i \sigma^x_i + \sum_{i,j} J_{ij} \sigma^z_i \sigma^z_j + \sum_{i,j} K_{ij} \sigma^x_i \sigma^x_j 
\]
where  coefficients $d_i, h_i, K_{ij}, J_{ij}$ are real numbers.

This Hamiltonian represents a 2-local Ising model with 1-local transverse field and a tunable 2-local transverse $\sigma^x\sigma^x$ coupling. The $\sigma^x\sigma^x$ is realisable, e.g., using capacitive coupling of flux qubits or with polar molecules \cite{BL08}.

\probTheorem{}{Biamonte and Love}{BL08}\\ 
\probRed{\langfont{2-local real Hamiltonian}}{\ref{realHam}}\\
\probClassical{When when $K_{ij}=d_i=0$ we obtain the famous Ising (spin glass) model with a magnetic field, which is NP-complete on a planar graph \cite{Barahona82}}
\\
\item \label{BL08-2} The 2-local Hamiltonian
\[
H_{ZX} = \sum_i h_i \sigma^z_i + \sum_i d_i \sigma^x_i + \sum_{i<j} J_{ij} \sigma^z_i \sigma^x_j + \sum_{i<j} K_{ij} \sigma^x_i \sigma^z_j 
\]
where  coefficients $d_i, h_i, K_{ij}, J_{ij}$ are real numbers.
The $\sigma^x\sigma^z$ is realisable using flux qubits \cite{BL08}.

\probTheorem{}{Biamonte and Love}{BL08} \\ 
\probRed{\langfont{2-local real Hamiltonian}}{\ref{realHam}}
\\
\item \label{Heisenberg} The 2D Heisenberg Hamiltonian with local magnetic fields

The 2D Heisenberg Hamiltonian is a model for spins on a 2-dimensional lattice in a magnetic system, and is often used to study phase transitions. It takes the form
\[
H_{\text{Heis}} = J\sum_{\langle i,j\rangle} \boldsymbol{\sigma_i}\cdot\boldsymbol{\sigma_j} - \sum_i \boldsymbol{\sigma_i}\cdot\boldsymbol{B}_i .
\]
Here, sums over $i$ range over all sites $i$ in the lattice, and $\langle i,j \rangle$ range over nearest-neighbouring sites. The local magnetic field at site $i$ is denoted by $\boldsymbol{B_i}$, and the coupling-constant $J$ is a real constant.
Hamiltonians restricted to this form are QMA-complete both for $J>0$ and for $J<0$.

\probTheorem{}{Schuch and Verstraete}{SV09}\\
\probRed{2-local 2D-lattice Hamiltonian}{\ref{2D-lattice}}
\\
\item \label{Hubbard} The 2D Hubbard Hamiltonian with local magnetic fields

The 2D Hubbard model describes a system of fermions on a 2-dimensional lattice and
is therefore used to model electrons in solid-state systems. 
It takes the form
\[
H_{\text{Hubb}} = -t\sum_{\langle i,j\rangle, s} a^\dag_{i,s} a_{j,s} + U\sum_i n_{i,\uparrow} n_{i,\downarrow} - \sum_i \boldsymbol{\bar \sigma_i}\cdot\boldsymbol{B}_i \ .
\]
Here, sums over $i$ range over all sites $i$ in the lattice, $\langle i,j \rangle$ range over nearest-neighbouring sites, and $s$ range over spins ${\lbrace \uparrow,\downarrow \rbrace}$. In this model, $\boldsymbol{\bar \sigma_i}$ is the Pauli matrices vector converted into orbital pair operators: $\boldsymbol{\bar \sigma_i}= {\lbrace \bar \sigma^x_i, \bar \sigma^y_i, \bar \sigma^z_i \rbrace}$ with $\bar \sigma^\alpha_i = \sum_{s,s'} \sigma^\alpha_{ss'} a^\dag_{i,s} a_{i,s'}$ where $\sigma^\alpha_{ss'}$ denotes the $(s,s')$ element of Pauli matrix $\sigma^\alpha$. The local magnetic field at site $i$ is denoted by $\boldsymbol{B_i}$, and $U$ and $t$ are positive numbers representing the electron-electron Coulomb repulsion and electron tunneling rate, respectively.

\probTheorem{}{Schuch and Verstraete}{SV09}\\
\probRed{Heisenberg Hamiltonian}{\ref{Heisenberg}} 

\end{enumerate}

\EnumItem{Translationally invariant $k$-local Hamiltonian}
\probIntro{There has been some interest in studying the \langfont{$k$-local Hamiltonian} (\ref{k-local}) problem with the added restriction that the Hamiltonian be \textit{translationally invariant}, i.e. that the Hamiltonian be identical at each particle (qudit\seeglos) in the system. Such systems are generally in a one-dimensional geometry with periodic boundary conditions. Some problems additionally employ geometric locality (which we refer to here as being on a line), such as constraining interactions to be between nearest-neighbouring particles, or between nearby (but not necessarily nearest-neighbouring) particles; some problems do not, however, have such geometric locality constraints.  Current results are listed here. These results are all built on Ref.\cite{AGIK09}. Finally, note that there may be complications in discussing QMA-completeness, since if a Hamiltonian is local and translationally invariant, the only input that scales is the number of qudits, $n$; it may need to be assumed that $n$ is given to the problem in unary to avoid these complications, but we will not discuss this here.
}

The \langfont{$k$-local Hamiltonian} problem is:
\begin{enumerate}
\item \probTheorem{even with a translationally invariant 3-local Hamiltonian with 22-state qudits, but where the interactions are not necessarily geometrically local.}{Vollbrecht and Cirac}{VC08}
\\
\item \probTheorem{even for translationally invariant 2-local Hamiltonians on $\poly(n)$-state qudits}{Kay}{Kay07}
\\
\item \probTheorem{even for translationally invariant $O(\log n)$-local Hamiltonians on 7-state qudits, where the interactions are geometrically local (albeit not restricted to nearest-neighbours)}{Kay}{Kay07}
\\
\item \probTheorem{even for 2-local Hamiltonians on a line of 49-state qudits where all strictly-2-local Hamiltonian terms are translationally invariant, although the 1-local terms can still be position-dependent}{Kay}{Kay08} 
\\
\probNotes{Although not discussed here, similar results exist for rotationally invariant Hamiltonians \cite{Kay09}.\\
There exist translationally invariant 2-local Hamiltonian problems on constant\hyp{}dimensional qudits, where the interactions are only between nearest-neighbours 
(and in which the only input is the size of the system) that are QMA$_{\text{EXP}}$-complete, where QMA$_{\text{EXP}}$ is the quantum analogue of the classical complexity class NEXP; see \cite{GI10}.}

\end{enumerate}

\EnumItem{Universal functional of DFT}
\probPreamble{
In quantum chemistry, \textit{density functional theory} (DFT) is a method for approximating the ground-state energy of an electron system (see \cite{SV09} and the references therein). The Hamiltonian for a system of $N$ electrons is $H = T^e + V^{ee} + V^e$ where the kinetic energy, electron-electron Coulomb potential, and local potential are given respectively by 
\[
\begin{split}
&  T^e = -\frac{1}{2}\sum_{i=1}^N \nabla^2_i \\
&  V^{ee} = \sum_{1\leqslant i < j \leqslant N} \frac{\gamma}{|\boldsymbol r_i - \boldsymbol r_j|} \\
&  V^e = \sum_{i=1}^N V(\boldsymbol x_i)
\end{split}
\]
where $\gamma>0$, $\boldsymbol r_i$ is the position of the $i$th electron, $\boldsymbol x_i = (\boldsymbol r_i, s_i)$ is the position $(\boldsymbol r_i$) of the $i$th electron together with its spin ($s_i$), and $\nabla^2$ is the Laplacian operator.

The ground-state energy of a system of $N$ electrons can be found by minimizing the energy over all $N$-electron densities $\rho^{(N)}(\boldsymbol{x})$, but it can also be given by minimizing over all single-electron probability distributions $n(\boldsymbol{x})$ as
\[ 
E_0 = \min_n \Big(\tr{V^e n(\boldsymbol{x})} + F[n(\boldsymbol{x})] \Big)
\]
where the \textit{universal functional of DFT} is 
\[
F[n(\boldsymbol{x})] = \min_{\rho^{(N)} \rightarrow n} \tr{(T^e + V^{ee})\rho^{(N)}(\boldsymbol{x})} .
\]
In the universal functional, the minimization is over all $N$-electron densities $\rho^{(N)}(\boldsymbol{x})$ that give rise to the reduced-density $n(\boldsymbol{x})$; therefore $F[n]$ gives the lowest energy of $T^e + V^{ee}$ consistent with $n$.  The difficult part of DFT is approximating $F[n(\boldsymbol{x})]$, which is independent of the external potential $V^e$ and is therefore universal for all systems.
}

\probDesc{
Given an integer $N$, representing the number of electrons, and a one-electron probability density $n(\boldsymbol{x})$}
{$F[n(\boldsymbol{x})] \leqslant a$}
{$F[n(\boldsymbol{x})] \geqslant b$}
{\wherePoly{b-a}{N} and the strength of the Hamiltonian is bounded by $\poly(N)$}

\probTheorem{}{Schuch and Verstraete}{SV09, WLA12}\\ 
\probRed{Hubbard model}{\ref{Hubbard}} [Turing reduction]

\end{enumerate} 
%
%
\subsection{Quantum $k$-SAT and its variations}
\langfont{Quantum $k$-SAT} is really just the \langfont{$k$-local Hamiltonian} problem restricted to projection operators. Nonetheless, it is included here as a subsection of its own due to its high level of interest and study. Note that occasionally people speak of the problem \langfont{MAX-quantum-$k$-SAT}; this is just another name for the \langfont{$k$-local Hamiltonian} problem (\ref{k-local}), and is therefore QMA-complete for $k\geqslant 2$. The problem \langfont{quantum $k$-SAT}, however, is different.\\

\setEnumLabel{S} \begin{enumerate}

\EnumItem{quantum $k$-SAT \label{q-k-SAT}}
\probIntro{
\langfont{Quantum $k$-SAT} is the quantum analogue of the classical problem $k$-SAT. It is actually simply the \langfont{$k$-local Hamiltonian problem} restricted to the case of $k$-local projector Hamiltonians\seeglos.
In classical $k$-SAT, the objective is to determine whether there exists a bit string (so each character in the string can be either 0 or 1) that satisfies (all of) a set of Boolean clauses, each of which only involves at most $k$ bits of the string. In the quantum analogue, rather than Boolean clauses we have projection operators. A \langfont{quantum $k$-SAT} instance has a solution if there is a quantum state that passes (i.e., is a 0-eigenvalue of) each projection operator.

We provide two equivalent definitions of this problem here. The first emphasises \langfont{quantum $k$-SAT} as a special case of \langfont{$k$-local Hamiltonian}, and the second emphasises the similarity to classical $k$-SAT.
}

\probDesc{
Given $k$-local projection operators $\lbrace \Pi_1,\ldots,\Pi_m \rbrace$ on $n$ qubits, where $m=\poly(n)$, and letting $H=\sum_{i=1}^m \Pi_i$}
{$H$ has an eigenvalue of precisely 0}
{all of the eigenvalues of $H$ are larger than $b$}
{\wherePoly{b}{n}}

Equivalently, we can define the problem as follows.\\
\probDesc{
Given polynomially many $k$-local projection operators $\lbrace\Pi_i\rbrace$}
{$\exists\ \ket{\psi}$ such that $\Pi_i \ket{\psi} = 0$ $\forall i$}
{$\forall\ \ketpsi, \sum_i \bra{\psi} \Pi_i \ket{\psi} \geqslant \epsilon $ (i.e. the expected number of `clause violations' is $\geqslant\epsilon$)} 
{\wherePoly{\epsilon}{n}}

\probTheoremO{for $k\geqslant 3$}{Gosset and Nagaj\footnote{improving the results of Bravyi\cite{Bravyi06}, which showed this for $k\geqslant 4$}}{GN13}\\ 
\probRed{QCSAT}{\ref{QCSAT}}

\probNotes{
\langfont{Quantum $k$-SAT} is in P for $k=2$.\cite{Bravyi06}\\
\langfont{Quantum $k$-SAT} is still QMA$_1$-complete if instead of demanding that $\Pi_i$ be projectors, we demand they be positive-semidefinite operators with zero ground-state energies and constant norms \cite{NM07}.
}

\probClassical{$k$-SAT is NP-complete for $k\geqslant 3$}

\EnumItem{quantum ($d_1,d_2,\ldots,d_k$)-SAT}
\probIntro{\langfont{Quantum ($d_1,d_2,\ldots,d_k$)-SAT} is a quantum $k$-SAT problem but with qudits rather than qubits. Specifically, in a \langfont{quantum ($d_1,d_2,\ldots,d_k$)-SAT} instance, the system consists of $n$ qudits (of various dimension), and each projection operator acts non-trivially on at most $k$ of these $n$ qudits, of the types specified, namely one $d_1$-dimensional qudit, one $d_2$-dimensional qudit,\ldots, and one $d_k$-dimensional qudit. Bear in mind that, e.g., if $d_1\geqslant d_2$ then a $d_2$-dimensional qudit is itself considered a type of $d_1$-dimensional qudit, so a projection operator that acts only on two $d_2$-dimensional qudits is also allowed. For example, an instance of \langfont{quantum (5,3)-SAT} involves a system of $n$ qudits (3-dimensional qudits called qutrits and 5-dimensional qudits called cinquits) such that each projection operator acts (non-trivially) on a single qudit, on one cinquit and one qutrit, or on two qutrits (but not on two cinquits).
For purposes of notation, we assume that $d_1\geqslant d_2 \geqslant \ldots \geqslant d_k$.

For $k\geqslant 3$, this class is trivial\footnote{Earlier work by Nagaj and Mozes\cite{NM07} that \langfont{quantum (3,2,2)-SAT} is QMA$_1$-complete is now subsumed by the result that \langfont{quantum 3-SAT} is QMA$_1$-complete.}: since \langfont{quantum 3-SAT} is QMA$_1$-complete for qubits, it is certainly QMA$_1$-complete for qudits. The cases of $k=2$ is not fully understood; however, the following results are known.}

\begin{enumerate}
\item 
\langfont{quantum (5,3)-SAT}, i.e. with a cinquit and a qutrit \\
\probTheoremO{for $k=2$ with $d_1\geqslant 5$, $d_2\geqslant 3$}{Eldar and Regev}{ER08}
\\
\item
\langfont{quantum (11,11)-SAT on a (one-dimensional) line} \\
\probTheoremO{}{Nagaj}{Nagaj08}\\
\end{enumerate}

\probNotes{
\langfont{Quantum (2,2)-SAT}, i.e. \langfont{quantum 2-SAT}, is in P.\\
\langfont{Quantum ($d_1,d_2$)-SAT} when $d_1<5$ or $d_2=2$ are open questions. They are known to be NP-hard (except for $d_1=d_2=2$ which is in P).\cite{ER08}
}

\probClassical{even though classical 2-SAT is in P, classical (3,2)-SAT, where one of the binary variables is replaced by a ternary variable, is NP-complete\cite{ER08}} 

\EnumItem{stochastic $k$-SAT}
\probIntro{This problem is like \langfont{quantum $k$-SAT}, except that instead of projection operators it uses stochastic, Hermitian, positive-semidefinite operators (see glossary, appendix~\ref{glossary}, for definitions).}

\probDesc{
Given polynomially many $k$-local stochastic, Hermitian, positive-semidefinite operators $\lbrace H_1,\ldots,H_m \rbrace$ on $n$-qubits with norms bounded by $\poly(n)$}
{the lowest eigenvalue of $H=\sum_i H_i$ is 0}
{all eigenvalues of $H$ are $\geqslant b$}
{\wherePoly{b}{n}}

\probTheoremO{for $k=6$}{Jordan, Gosset, and Love}{JGL10}\\
\probRed{\langfont{quantum 4-SAT}}{\ref{q-k-SAT}}

\probNotes{
\langfont{Stoquastic quantum $k$-SAT}, where the word `stochastic' is replaced by `stoquastic' above, is in MA and is MA-complete for $k\geqslant 6$. \cite{BT08} Note that \langfont{stochastic $k$-SAT} makes no mention of projection operators, and therefore isn't really a quantum $k$-SAT problem. In \langfont{stoquastic quantum $k$-SAT}, its MA-complete cousin, however, the operators can be converted to equivalent operators that are projectors, whence the relation to \langfont{quantum $k$-SAT}. No connection to projectors is known in the stochastic case.
}

\end{enumerate} 
%
%
\section{Density matrix consistency}
\setEnumLabel{C} \begin{enumerate}

\EnumItem{$k$-local density matrix consistency \label{local_consistency}}
\probIntro{
Given a set of density matrices on subsystems of a constant number of qubits, this problem is to determine whether there is a global density matrix on the entire space that is consistent with the subsystem density matrices.
}


\probDesc{
Consider a system of $n$ qubits. Given $m=\poly(n)$ \quad $k$-local density matrices $\rho_1, \ldots, \rho_m$, so that each $\rho_i$ acts only on a subset $C_i \subseteq \lbrace 1,\ldots n \rbrace$ of qubits  with $|C_i| \leqslant k$}
{$\exists$ $n$-qubit density matrix $\sigma$ such that $\forall i$, $\Vert\rho_i - \tilde\sigma_i\Vert_{tr} = 0$ where $\tilde \sigma_i = \trP_{\{1,\ldots, n\}\smallsetminus C_i} (\sigma)$}
{$\forall$ $n$-qubit density matrix $\sigma$, $\exists i$ such that $\Vert\rho_i - \tilde\sigma_i\Vert_{tr} \geqslant b$ where $\tilde \sigma_i = \trP_{\{1,\ldots, n\}\smallsetminus C_i} (\sigma)$}
{\wherePoly{b}{n}}

\probTheorem{even for $k=2$}{Liu}{Liu06}\\
\probRed{\langfont{$k$-local Hamiltonian}}{\ref{k-local}} [Turing reduction] 

\probClassical{\langfont{consistency of marginal distributions} is NP-hard}

\EnumItem{$N$-representability}
\probIntro{
This is the same problem as \langfont{$2$-local density matrix consistency} (\ref{local_consistency}), but specialised to fermions (particles whose quantum state must be antisymmetric under interchange of particles).
}

\probDesc{
Given a system of $N$ fermions and $d$ possible modes, with $N\leqslant d\leqslant \poly(N)$, and a $\frac{d(d-1)}{2} \times \frac{d(d-1)}{2}$ 2-fermion density matrix $\rho$}
{$\exists$ ${d \choose N} \times {d \choose N}$  $N$-fermion density matrix $\sigma$ such that $\trP_{\lbrace 3,\ldots,N\rbrace}(\sigma) = \rho$}
{$\forall$ $N$-fermion density matrices $\sigma$, $\Vert\rho - \trP_{\lbrace 3,\ldots,N\rbrace}(\sigma)\Vert_{tr} \geqslant b$}
{\wherePoly{b}{N}}

\probTheorem{}{Liu, Christandl, and Verstraete}{LCF07}\\
\probRed{\langfont{2-local Hamiltonian}}{\ref{k-local}} [Turing reduction]


\EnumItem{bosonic $N$-representability}
\probIntro{
This is the same problem as \langfont{$2$-local density matrix consistency} (\ref{local_consistency}), but specialised to bosons (particles whose quantum state must be symmetric under interchange of particles).
}

\probDesc{
Given a system of $N$ bosons and $d$ possible modes, with $d\geqslant cN$ (for some constant $c>0$), and a $\frac{d(d+1)}{2} \times \frac{d(d+1)}{2}$ 2-boson density matrix $\rho$}
{$\exists$ ${N+d-1 \choose N} \times {N+d-1 \choose N}$  $N$-boson density matrix $\sigma$ such that $\trP_{\lbrace 3,\ldots,N\rbrace}(\sigma) = \rho$}
{$\forall$ $N$-boson density matrices $\sigma$, $\Vert\rho - \trP_{\lbrace 3,\ldots,N\rbrace}(\sigma)\Vert_{tr} \geqslant b$}
{\wherePoly{b}{N}}

\probTheorem{}{Wei, Mosca, and Nayak}{WMN10} \\
\probRed{2-local Hamiltonian}{\ref{k-local}} [Turing reduction]

\end{enumerate} 

%
%
\appendix
%
%
\section{Glossary } 
\label{glossary}
The definitions given here are not necessarily the most general or precise possible, but they suffice for the needs of this paper.

\newcommand{\GlosItem}[1]{\item[#1] --}
\begin{description}

\GlosItem{$i^\text{th}$ energy of a Hamiltonian $H$} the $i^\text{th}$ smallest eigenvalue of $H$.

\GlosItem{ground-state energy of a Hamiltonian $H$} the smallest eigenvalue of $H$.

\GlosItem{Hamiltonian} the generator of time-evolution in a quantum system. Its eigenvalues correspond to the allowable energies of the system. It also dictates what interactions are present in a system. As a matrix, it is Hermitian.

\GlosItem{Hermitian matrix} a square matrix $H$ that is equal to its own conjugate-transpose, i.e. $H^\dag = H$.

\GlosItem{norms of matrices} Several different matrix norms appear in this paper. 
Given a matrix $A$ with elements $a_{ij}$, the
\begin{description}

\item[operator norm] of $A$ is 
$\Vert A\Vert = \max{\left\lbrace      \Vert A\ketpsi\Vert_2 : \Vert\ketpsi\Vert_2 =1      \right\rbrace}$. For a square matrix, it is also known as the spectral norm; it is the largest singular value of $A$, and if $A$ is normal, then it is the largest absolute value of the eigenvalues of $A$.

\item[Frobenius norm] of $A$ is $\Vert A\Vert_F = \sqrt{\tr{A^\dag A}} = \sqrt{\sum_{i,j} |a_{ij}|^2}$. 

\item[trace norm] of $A$ is $\Vert A\Vert_{tr} = \tr{\sqrt{A^\dag A}}$, 
which when $A$ is normal is the sum of the absolute value of its eigenvalues. It is often written $\Vert A\Vert_{tr} = \trP|A|$ where $|A|$ denotes $\sqrt{A^\dag A}$.

\end{description}

\GlosItem{norms of quantum superoperators} Occasionally norms of superoperators are required in this paper.
\begin{description}
\item[diamond norm] of a superoperator $\Phi$ that acts on density matrices that act on a Hilbert space $\cal{H}$ is $\Vert \Phi \Vert_\diamondsuit = \sup_{X}\Vert(\Phi\otimes\id)(X)\Vert_{tr}/\Vert X\Vert_{tr}$ where the supremum is taken over all linear operators  
 $X:\cal{H}\otimes \cal{H} \rightarrow \cal{H}\otimes \cal{H}$.
\end{description}

\GlosItem{positive-semidefinite matrix}
a Hermitian matrix whose eigenvalues are all non-negative.

\GlosItem{$k$-local projector on $n$ qubits} a Hermitian matrix of the form $\Pi = \id^{\otimes (n-k)}\otimes \sum_i\selfketbra{\psi}_i$ where the $\ket{\psi}_i$ are orthonormal $k$-qubit states. It satisfies $\Pi^2 = \Pi$. 

\GlosItem{stochastic matrix}
a square matrix of non-negative real numbers such that each row sums to 1. If additionally each column sums to 1, it is called a doubly stochastic matrix.


\GlosItem{stoquastic Hamiltonian}
a Hamiltonian in which the off-diagonal matrix elements are non-positive real numbers in the standard basis.

\GlosItem{qudit}
generalization of a qubit: for some $d$, a $d$-state quantum-mechanical system, or mathematically, a unit-normalized vector in $\C^d$ (but where global phase is irrelevant). When $d=2$ it is called a qubit, when $d=3$ it is called a qutrit. When $d=5$ it may be called a cinquit\cite{ER08}, but to avoid headaches, I advise against trying to name the $d=4$ version.

\end{description}

%
%
\section{QMA-hard theorems \label{app:hard}}
This appendix contains a theorem that allows one to prove that several quantum circuit verification problems are QMA-hard. Note that it doesn't prove QMA-completeness, only QMA-hardness, so it is relegated to the appendix.

\setEnumLabel{X} \begin{enumerate}

\EnumItem{Quantum circuit testing \label{circuit_testing}}
%
\probIntro{This problem involves testing the behaviour of a quantum circuit. Given input circuit $C$, one wishes to determine whether it acts like a circuit from uniform circuit family $\mathscr{C}_0$ on a large input space, or like a circuit from uniform circuit family $\mathscr{C}_1$ for all inputs, promised that the two families are significantly different.}

\probDesc{
Let $\delta\in(0,1]$ and let $\mathscr{C}_0$ and $\mathscr{C}_1$ be two uniform families of quantum circuits. Given input quantum circuit $C$ acting on $n$-qubit input space ${\cal H}$, let $C_0\in \mathscr{C}_0$ and $C_1\in \mathscr{C}_1$ act on the same input space ${\cal H}$. The problem is}
{$\exists$ subspace $S$, with $\text{dim} S \geqslant (\text{dim} {\cal H})^{1-\delta}$, such that for any reference space ${\cal R}$, if $\rho$ is a density matrix on $S\otimes {\cal R}$ then $\Vert (C\otimes\id_{{\cal R}})(\rho) - (C_0\otimes\id_{\cal R})(\rho)\Vert_{tr} \leqslant \epsilon$}
{for any reference space ${\cal R}$, if $\rho$ is a density matrix on the full space ${\cal H}\otimes {\cal R}$ then $\Vert (C\otimes\id_{\cal R})(\rho) - (C_1\otimes\id_{\cal R})(\rho)\Vert_{tr} \leqslant \epsilon$}
{where $1 > \epsilon \geqslant 2^{-\poly(n)}$. Note that the promise actually imposes a condition on the allowable $\mathscr{C}_0$ and $\mathscr{C}_1$, forcing them to be significantly different}

\probTheoremH{for constant $\delta$}{Rosgen}{Rosgen11b}\\   
\probRed{QCSAT}{\ref{QCSAT}}

\probNotes{leads to: \langfont{mixed-state non-identity check} (\ref{mixed-non-identity}), \langfont{non-isometry testing} (\ref{non-isometry}), \langfont{Detecting insecure quantum encryption}(\ref{insecure_q_enc})}

\end{enumerate}  

%
%
\section{Diagram of QMA-complete problems}
\begin{figure}[h] 
	\begin{center}
		\includegraphics[height=4in]{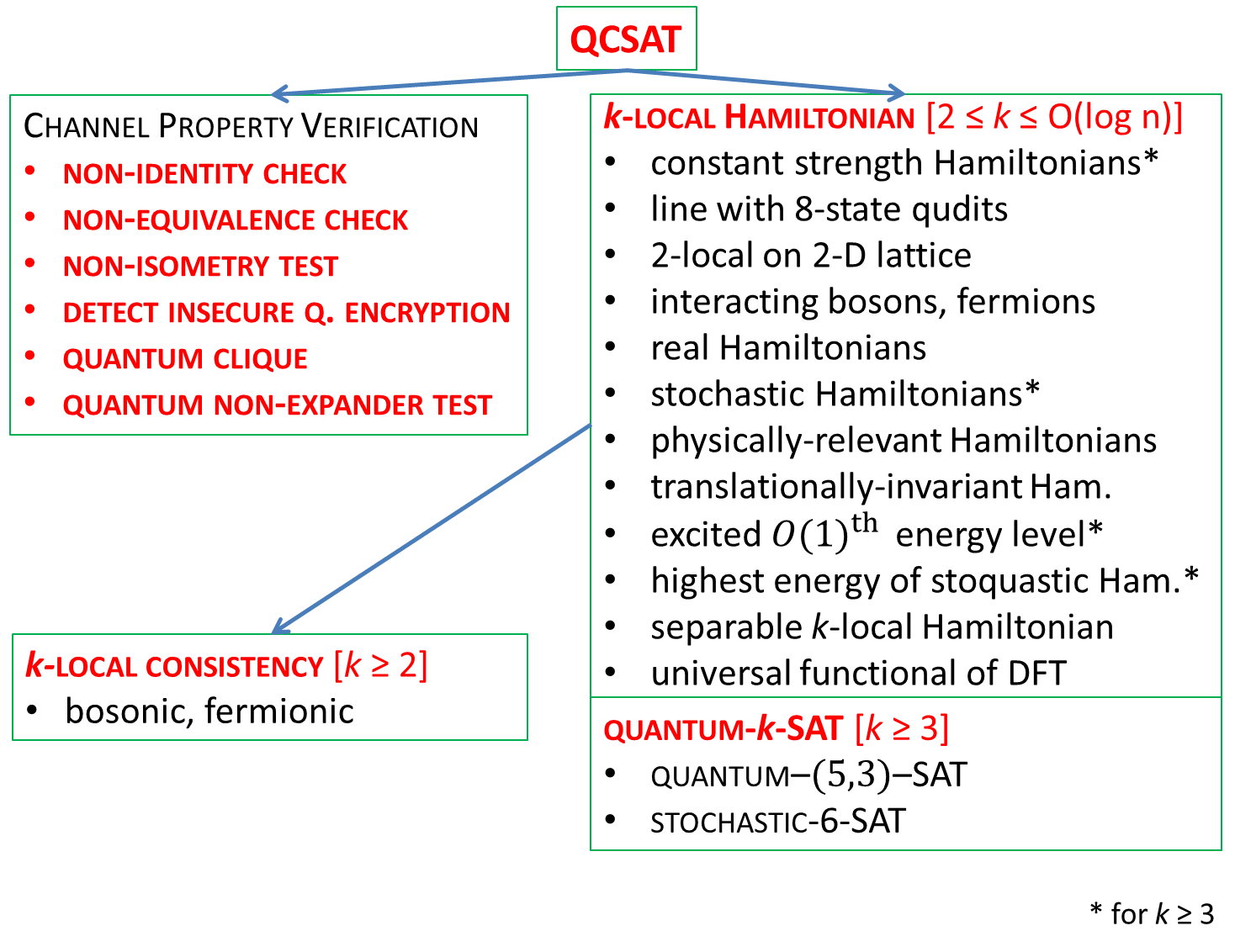}
		\caption{\label{fig:tree}Schematic showing the QMA-complete problems listed in this paper, according to their categories. Lines show hardness reductions.}
	\end{center}
\end{figure}

%
%

\subsection*{Acknowledgments}
The author gratefully acknowledges support from the Department of Physics at MIT. Much appreciation goes to Scott Aaronson, for whose excellent MIT 6.845 course I prepared this paper, and to Pawel Wocjan, for greatly advancing my knowledge of the class QMA and its history.

%
%

\end{document}